\documentclass[a4paper]{aa}

\usepackage{txfonts}
\usepackage[english]{babel}
\usepackage[dvips]{graphicx}
\usepackage{latexsym}
\usepackage{amssymb}
\usepackage{mathrsfs}

\title{Large-scale structure of Lyman break galaxies \\ around a radio galaxy protocluster at $z \sim 4$\thanks{Based on data collected at Subaru Telescope, which is operated by the National Astronomical Observatory of Japan.}}

\author{
H.T.~Intema \inst{1} \and
B.P.~Venemans \inst{1} \and
J.D.~Kurk \inst{1,2} \and
M.~Ouchi \inst{3,4} \and \\
T.~Kodama \inst{5,6} \and
H.J.A.~R\"ottgering \inst{1} \and
G.K.~Miley \inst{1} \and
R.A.~Overzier \inst{1}
}

\institute{
Sterrewacht Leiden, PO Box 9513, NL-2300 RA, Leiden, The Netherlands \\
\email{intema@strw.leidenuniv.nl} \and
INAF, Osservatorio Astrofisico di Arcetri, Largo E. Fermi 5, 50125, Firenze, Italy \and
Space Telescope Science Institute, 3700 San Martin Drive, Baltimore, MD 21218, USA \and
Hubble Fellow \and
National Astronomical Observatory of Japan, Mitaka, Tokyo 181-8588, Japan \and
European Southern Observatory, Karl-Schwarzschild-Str. 2, D-85748 Garching, Germany
}

\date{Received January 1, 2006 / Accepted June 7, 2006}

\begin{document}

\newcommand{\as}{arcsec}
\newcommand{\am}{arcmin}
\newcommand{\dg}{deg}
\newcommand{\mpc}{Mpc}
\hyphenation{arc-sec arc-min}

\abstract{
We present broad-band imaging with the Subaru Telescope of a $25' \times 25'$ field surrounding the radio galaxy TN~J1338--1942 at redshift $z = 4.1$. The field contains excesses of Lyman-$\alpha$ emitters (LAEs) and Lyman break galaxies (LBGs) identified with a protocluster surrounding the radio galaxy. Our new wide-field images provide information about the boundary of the protocluster and its surroundings. There are 874~candidate LBGs within our field, having redshifts in the range $z = 3.5 - 4.5$. An examination of the brightest of these (with $i' < 25.0$) shows that the most prominent concentration coincides with the previously discovered protocluster. The diameter of this galaxy overdensity corresponds to $\sim 2$~\mpc{} at $z = 4$, consistent with the previous estimation using LAEs. Several other concentrations of LBGs are observed in the field, some of which may well be physically connected with the $z = 4.1$ protocluster. The observed structure in the smoothed LBG distribution can be explained as the projection of large-scale structure, within the redshift range $z = 3.5 - 4.5$, comprising compact overdensities and prominent larger voids. If the $5 - 8$ observed compact overdensities are associated with protoclusters, the observed protocluster volume density is $\sim 5 \times 10^{-6}$~\mpc$^{-3}$, similar to the volume density of rich clusters in the local Universe.
}

\maketitle

\section{Introduction}
\label{sec:int}

There is considerable evidence for galaxy overdensities at high redshifts ($z > 2$; e.g. Steidel et al. \cite{bib:steidel1998}; M\o{}ller \& Fynbo \cite{bib:moller2001}; Shimasaku et al. \cite{bib:shimasaku2003}; Palunas et al. \cite{bib:palunas2004}; Ouchi et al. \cite{bib:ouchi2005}). In many cases these overdensities have been presumed to be associated with the ancestors of rich local clusters. At $z > 2$ the Universe is $\lesssim 3$~Gyr old, too short for these structures to have virialized (e.g. Venemans \cite{bib:venemans2005}). Hence these structures are often called \emph{protoclusters}. Most searches for protoclusters have been limited by relatively small fields (typically smaller than $10' \times 10'$).

An efficient way of finding protoclusters is to use high redshift radio galaxies (HzRGs; e.g. R\"ottgering et al. \cite{bib:rottgering1994}) as tracers (Venemans et al. \cite{bib:venemans2002}; Kurk et al. \cite{bib:kurk2004}). HzRGs are large massive objects with many of the properties expected of forming dominant cluster (cD) galaxies (West \cite{bib:west1994}). Although most protoclusters are not radio-loud, radio-selected protoclusters may be typical. Because radio-sources are relatively short-lived ($\sim 10^7 $ years; Blundell \& Rawlings \cite{bib:blundell1999}), the statistics are consistent with the progenitor of every rich local cluster having harboured a luminous radio galaxy at some stage in its existence.   

Using the VLT, Venemans et al. (\cite{bib:venemans2002}) spectroscopically confirmed 20 Lyman-$\alpha$ emitters (LAEs) in a $7' \times 7'$ field around HzRG TN~J1338--1942 at a redshift of $z = 4.1$ and identified these LAEs with a $z = 4.1$ protocluster. Further evidence that this LAE overdensity was indeed associated with a protocluster was provided by observations with the HST/ACS that revealed an excess and non-uniform distribution of candidate Lyman break galaxies (LBGs) around the radio galaxy (Miley et al. \cite{bib:miley2004}; Overzier et al. \cite{bib:overzier2005}).

Although the LAE search around TN~J1338--1942 was extended with a second $7' \times 7'$ field (Venemans \cite{bib:venemans2005}), this was insufficient to determine the boundary of the protocluster. Here we present the results of a multi-color study of candidate LBGs from a $25' \times 25'$ region surrounding TN~J1338--1942. The large field-of-view (FOV) facilitates searches for LBGs out to the boundary of the protocluster structure and beyond. The data also provide new information about large-scale structure and voids at $z \sim 4$. 

Throughout this paper, we use AB-magnitudes, $1 \sigma$ errors and adopt a flat, $\Lambda$-dominated cosmology with $\Omega_\mathrm{M} = 0.3$, $\Omega_{\Lambda} = 0.7$ and $H_0 = 100~h$~km~s$^{-1}$~\mpc$^{-1}$ with $h = 0.7$.

\section{Data reduction and sample selection}
\label{sec:drss}

Deep multi-color imaging of the TN~J1338--1942 field was carried out using the Subaru/Suprime-Cam instrument (Miyazaki \cite{bib:miyazaki2002}) on January $31^{\mathrm{st}}$ and February $1^{\mathrm{st}}$, 2003. Data reduction on the $B$-, $R_\mathrm{C}$- and $i'$-band images was performed in a manner similar to that of the SDF and SXDF fields (Ouchi et al. \cite{bib:ouchi2001}; \cite{bib:ouchi2004a}). The FOV was $24.7' \times 24.2'$ and the seeing had an equivalent FWHM of $0.98''$. Source extraction and photometry was done using SExtractor (Bertin \& Arnouts \cite{bib:bertin1996}). Fluxes were measured in a $2''$ circular aperture with $3 \sigma$ limiting magnitudes of $27.2$, $26.9$ and $26.5$, respectively. Extending the observed power-law estimation for bright source counts (e.g. see Palunas et al. \cite{bib:palunas2004}), the initial object sample was limited to $i' \leq 26.5$ to obtain a photometric completeness of $\sim 72\%$ for the highest magnitude bin ($\Delta i' = 0.5$).

LBGs in an approximate redshift range $z = 3.5 - 4.5$ were selected using color selection criteria by Ouchi et al. (\cite{bib:ouchi2004a}), resulting in a sample of 874~LBGs. Based on the same work by Ouchi et al., the estimated contamination by interlopers and stars is $\sim 6\%$, while the estimated completeness distribution function over redshift has an approximate gaussian function shape with a FWHM of $0.8$, centered at $z = 4$ with a peak value of $\sim 45\%$. In addition, a bright sub-sample of LBGs was constructed having $i' < 25.0$. For this sub-sample of 125 objects, the galaxy colors at the corresponding LBG redshift range are well constrained. The completeness distribution function for this sub-sample has a FWHM of 0.8, centered at $z = 4.1$ with a peak value of $\sim 90\%$, while the estimated contamination is $< 1\%$.

\section{Analysis}
\label{sec:lss}

\subsection{Projected density distribution of bright LBGs}
\label{sec:lss_pddb}

\begin{figure}
\resizebox{\hsize}{!}{\includegraphics[angle=90]{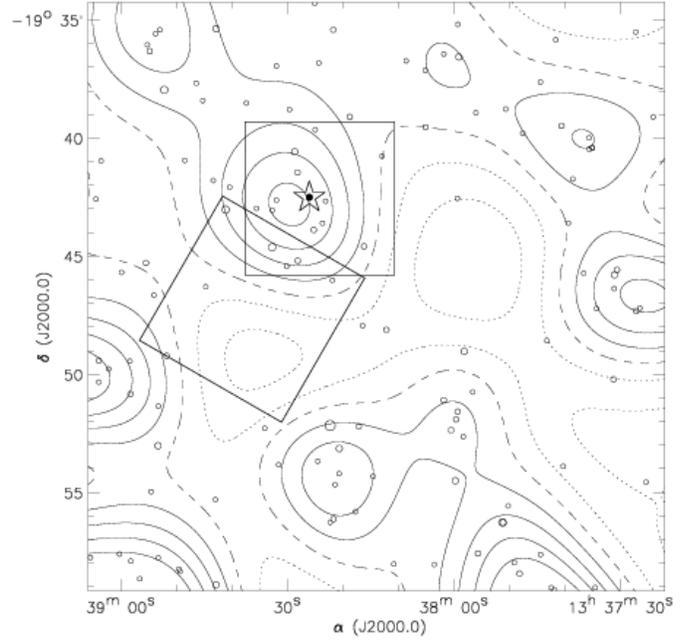}}
\caption{Projected distribution of 125~bright LBGs (open circles; diameter scales with brightness; $21.5 < i' < 25.0$) in the TN~J1338--1942 field, including TN~J1338--1942 (filled circle within star). The projected density contours (curved lines), obtained by gaussian smoothing, reveal overdense (solid lines; $\Delta = 0.25,~0.50,~0.75,~1.00$ from edge to center) and underdense regions (dotted lines; $\Delta = -0.25,~-0.50,~-0.75$ from edge to center) relative to a mean density of $0.21$ per square arcminute (dashed line). TN~J1338--1942 inhabits a significant overdense area, probably associated with the protocluster found by Venemans et al. (\cite{bib:venemans2002}). The rectangles represent the two fields that were used by Venemans et al. (\cite{bib:venemans2002}) and Venemans (\cite{bib:venemans2005}) to search for LAEs.}
\label{fig:lbg_map_125}
\end{figure}

We first investigated the projected distribution of the brightest LBGs by smoothing the spatial distribution of our bright sub-sample with a gaussian kernel. Structure identification is dependent on the size of the smoothing kernel. The FWHM of $5'$ was chosen to match the average distance between these LBGs, thereby optimizing the contrast between overdense and underdense regions. This FWHM is also similar to the angular size of LAE proto-clusters at $z = 3 - 6$ (Shimasaku et al. \cite{bib:shimasaku2003}; Ouchi et al. \cite{bib:ouchi2005}; Venemans et al. \cite{bib:venemans2005}), which improves the chance of detecting such structures in the LBG distribution. The smoothed LBG map was divided by a normalisation map to correct for undetected LBGs behind foreground objects and FOV boundaries, which causes some dense areas near the border to be overemphasized. The result is shown in Fig.~\ref{fig:lbg_map_125}.

The most significant projected overdensity in this bright LBG structure map (away from the border) contains the radio galaxy TN~J1338--1942. We associate this overdensity with the protocluster at $z = 4.1$ previously discovered in the smaller fields accessed by the VLT (LAEs in two $7' \times 7'$ fields; Venemans et al. \cite{bib:venemans2002}; Venemans \cite{bib:venemans2005}) and ACS/HST (LBGs in a $3' \times 3'$ field; Miley et al. \cite{bib:miley2004}; Overzier et al. \cite{bib:overzier2005}). Taking $\Delta = 0.5$ as the boundary (with $\Delta = (\Sigma - \langle \Sigma \rangle ) / \langle \Sigma \rangle$, where $\Sigma$ and $\langle \Sigma \rangle$ are the local and average projected LBG densities, respectively), (i) the diameter of the overdensity is $\sim 5'$, corresponding to a (proper) size of $\sim 2$~\mpc{} at $z = 4$ and (ii) the location of the radio galaxy is in the western part of the structure. The protocluster size and the relative location of the radio galaxy are similar to those found by Venemans et al. for LAEs.

\subsection{LBG overdensity in redshift space}
\label{sec:lss_clu_ors}

We determined the overlap between the whole LBG sample and the LAE sample by Venemans et al. (\cite{bib:venemans2002}) and Venemans (\cite{bib:venemans2005}) to obtain spectroscopic redshifts for several LBGs. There were 86~LAE candidates found in a narrow redshift range $\Delta z = 4.087 - 4.119$ surrounding TN~J1338--1942 (fields plotted in Figs.~\ref{fig:lbg_map_125} and \ref{fig:lbg_map_874}). Spectroscopic confirmation of redshift followed for 38~LAEs, including TN~J1338--1942. 

Within the same area, we identified 104~candidate LBGs (out of the whole LBG sample). Of these LBGs, 7 are also spectroscopically confirmed LAEs, thus obtaining a redshift for these 7~LBGs. Based on the contamination fraction, we expect 6 of the 104~candidate LBGs to be interlopers. To see whether the 7~confirmed LBGs represent a significant overdensity, we estimated the expected number of LBGs in the redshift range $\Delta z$ using Monte-Carlo simulations.  We randomly assigned redshifts to 98~objects, using the completeness distribution function as redshift distribution, and counted the number of objects in $\Delta z$. We repeated this procedure 10,000 times and found that the expected number of LBGs in redshift range $\Delta z$ is $3.6 \pm 1.9$. As a result, the 7~spectroscopically confirmed LBGs might indicate a modest LBG volume overdensity of $\delta = 1.0 \pm 1.0$ (with $\delta = ( \rho - \langle \rho \rangle ) / \langle \rho \rangle$, where $\rho$ and $\langle \rho \rangle$ are the local and average LBG volume density, respectively) in close vicinity of TN~J1338--1942. 

The estimated overdensity above is a lower limit, because not all high redshift galaxies have Ly-$\alpha$ emission. Steidel et al. (\cite{bib:steidel2000}) found that at $z = 3$ only $20 - 25 \%$ of the LBGs satisfy typical LAE selection criteria. Assuming that this is also true at $z = 4$, this implies that the 7~confirmed LBGs represent a true number of at least $28$~LBGs within $\Delta z$. As these are expected to be part of the 104~observed LBGs, the LBG volume overdensity within $\Delta z$ is increased to at least $\delta = 7 \pm 4$. Note that still $\sim 70 \%$ of the LBGs lie outside $\Delta z$. This excess of LBGs in redshift space is consistent with the location of the protocluster that harbours TN~J1338--1942.

\subsection{Projected density distribution of all LBGs}
\label{sec:lss_pdd}

\begin{figure}
\resizebox{\hsize}{!}{\includegraphics[angle=90]{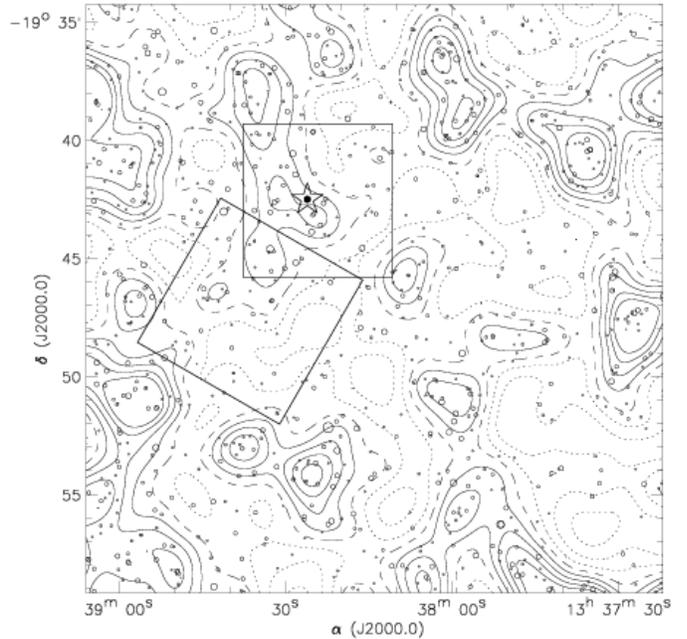}}
\caption{Projected distribution of 874~LBGs with $21.5 < i' < 26.5$ in the TN~J1338--1942 field, similar to Fig.~\ref{fig:lbg_map_125}. The contours are relative to a mean density of $1.47$ per square arcminute (dashed line). Like in Fig.~\ref{fig:lbg_map_125}, TN~J1338--1942 inhabits a significant overdense area, probably associated with the protocluster found by Venemans et al. (\cite{bib:venemans2002}).}
\label{fig:lbg_map_874}
\end{figure}

Using the same technique as in Sec.~\ref{sec:lss_pddb} (but using a FWHM of $2'$ to match the average distance between the LBGs), the whole sample of 874~LBGs was used to make a second structure map, which is shown in Fig.~\ref{fig:lbg_map_874}. In this map, TN~J1338--1942 also inhabits a clear but less prominent projected overdensity of galaxies. In addition to this overdensity associated with the previously known protocluster, $4 - 7$ other intriguing peaks are seen in the large-scale structure distribution, surrounded by larger regions of relatively empty space. These overdensities may be associated with protoclusters within the redshift range $z = 3.5 - 4.5$, while the underdense regions indicate the presence of large voids. The typical transverse size of the overdensities is $\sim 2$~\mpc{} at $z = 4$, while the underdensities are more than twice this size. Without spectroscopic data, we cannot establish whether the other overdensities are physically linked to the protocluster.

The overdensity surrounding TN~J1338--1942 contains relatively many bright LBGs compared to the other overdensities, which suggests that the TN~J1338--1942 protocluster is the most massive structure within the observed volume. The strong clustering of bright LBGs at one particular position within the FOV agrees with the observation that at $z \sim 4$, brighter LBGs have larger clustering lengths than fainter LBGs (Allen et al. \cite{bib:allen2005}; Ouchi et al. \cite{bib:ouchi2004b}).

Similar structure maps were created for 20~mock samples of 874~random points each with the same positional constraints as the LBG sample. Visual comparison between the detailed LBG map and the mock maps shows that for the latter, the overdensities are larger but lower in amplitude, while the underdensities are smaller and more isolated. Basically, these over- and underdensities have similar sizes and amplitudes, very different from what is observed in the LBG map.

\begin{figure}
\begin{center}
\resizebox{\hsize}{!}{\includegraphics[angle=90]{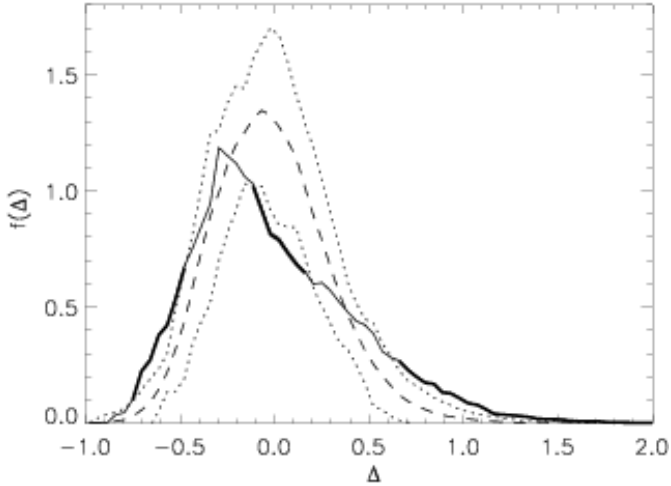}}
\caption{Plotted here are the (normalised) projected density distribution function (PDDF) of the whole sample of 874~LBGs (solid line) and the combined PDDF of 20~equally sized mock samples (dashed line is mean, dotted lines are mean $\pm~3 \sigma$, where $\sigma$ is the standard deviation between the mock samples). At several intervals, the PDDF of the LBG sample deviates from the mock samples by more than $3 \sigma$ (thick solid line). The smoothed LBG distribution has relatively less area with a density close to mean, relatively more area out to higher densities and relatively more area with lower densities, which characterizes the presence of compact overdensities with high peaks and extended underdensities.}
\label{fig:area_hist}
\end{center}
\end{figure}

Figure~\ref{fig:area_hist} shows the projected density distribution function (PDDF) of the detailed LBG map and the mean PDDF of the mock maps. Using a Kolmogorov-Smirnov test (e.g. Press et al. \cite{bib:press1992}), we found that the probability that the PDDF of the LBG map is drawn from an underlying distribution equal to the mean PDDF of the mock maps is $43\%$. For the 20~mock maps, the probability that they are drawn from the mean PDDF is much higher ($> 99.9\%$). In Figure~\ref{fig:area_hist}, it can be seen that for the LBG sample there is significantly ($> 3 \sigma$) more area with $-0.75 < \Delta < -0.5$ and $\Delta > 0.65$ than for the mock samples. This is consistent with the underdensities (presumably voids) being larger than the more strongly peaked overdensities (presumably protoclusters).

\subsection{LBG angular and spatial correlation}
\label{sec:lss_clu_asc}

The two-point angular correlation function (ACF; Peebles \cite{bib:peebles1973}; \cite{bib:peebles1980}) of the whole LBG sample was calculated, using estimators by Landy \& Szalay (\cite{bib:landy1993}) and Hamilton (\cite{bib:hamilton1993}), which gave practically identical results. The estimators are negatively offset from the true ACF due to the difference between the measured average and the true average LBG density in the restricted FOV (the `integral constraint'). When assuming a power-law form $\omega(\theta) = A_{\omega} \, \theta^{-\beta}$ for the true ACF, the offset can be estimated following Roche et al. (\cite{bib:roche2002}). Iteratively fitting a power-law and estimating the offset converged to solutions for both a variable ($\beta = 1.1 \pm 0.1$) and a fixed slope ($\beta = 0.8$; e.g. Peebles \cite{bib:peebles1980}) power-law. We found that for all cases the clustering amplitude $A_{\omega}$ is significantly larger than its uncertainty ($\geq 6 \sigma$), confirming that there is a non-random clustering signal present in our LBG distribution.

After correcting the clustering amplitude for contamination (Ouchi et al. \cite{bib:ouchi2004b}), the inverse Limber transformation (Efstathiou et al. \cite{bib:efstathiou1991}) was used to calculate the (comoving) spatial correlation length. We found correlation lengths of $r_0 = ( 3.7 \pm 0.7 )~h^{-1}$~\mpc{} and $r_0 = ( 4.6 \pm 0.4 )~h^{-1}$~\mpc{} for the variable and fixed slope fit, respectively. These results are similar to those found by Ouchi et al. (\cite{bib:ouchi2004b}), indicating that (within the observed volume) the clustering properties of LBGs in the radio galaxy field are not significantly different from blank fields.

\subsection{LBG void probability function}
\label{sec:lss_vpf}

\begin{figure}
\begin{center}
\resizebox{\hsize}{!}{\includegraphics[angle=90]{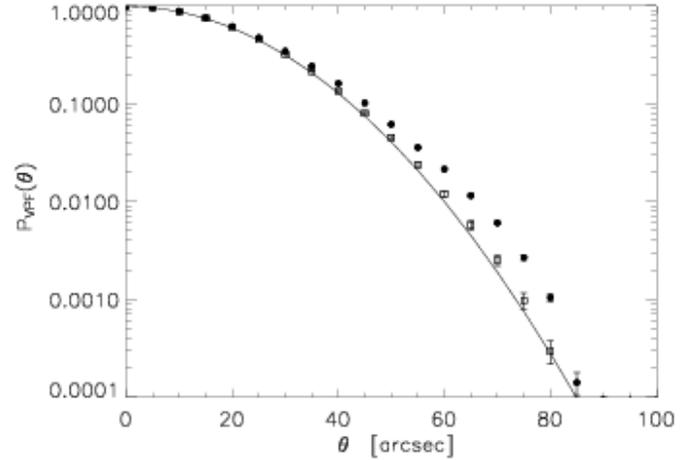}}
\caption{The two-dimensional void probability function (VPF) of the LBG sample (filled dots; error bar is poissonian error), compared with the mean VPF of 20~mock samples (open squares; error bar is combination of poissonian error and standard deviation between mock catalogs) and the theoretical VPF for poissonian distributions (solid line). For radii $\theta > 40''$, the underdensities in the projected LBG distribution are significantly larger than the underdensities in mock and poissonian distributions.}
\label{fig:vdf}
\end{center}
\end{figure}

A disadvantage of using the two-point ACF for detecting clustering in galaxy distributions is that it is less sensitive to non-gaussian density fluctuations. One can use $n$-point ACFs to improve the detection of the latter, but error-bars becomes exceedingly large with higher $n$ when the number of galaxies is fixed. A complementary approach is to focus on voids instead. The two-dimensional void probability function (VPF; White \cite{bib:white1979}) defines the fraction of circular areas at random positions in the FOV which contain no galaxies. The VPF was calculated for both the whole LBG sample and the 20~mock samples from Sec.~\ref{sec:lss_pdd} and plotted in Fig.~\ref{fig:vdf}. The mean VPF of the mock samples is very similar to the theoretical VPF for poissonian distributions (which is $P_\mathrm{VPF}(\theta) = \exp{ [ - \pi \theta^2 \langle \Sigma \rangle ] }$). At larger radii, the VPF of the full LBG sample is significantly higher than both the mean mock VPF and the theoretical VPF, meaning that the underdensities in the projected LBG distribution are relatively large compared to underdensities found in random distributions. This result is similar to the result of Palunas et al. (\cite{bib:palunas2004}) at $z = 2.34$, using a sample of 34~LAEs in a FOV similar to ours.

\subsection{Protocluster volume density}
\label{sec:dis}

The comoving volume that is observed using the LBG selection criteria by Ouchi et al. (\cite{bib:ouchi2004a}) has a size of approximately $1.8 \times 10^6$~\mpc{}$^3$. Assuming that $5 - 8$ observed overdensities in the LBG map are indeed associated with protoclusters, the volume density of protoclusters within $z = 3.5 - 4.5$ is $\sim 5 \times 10^{-6}$~\mpc$^{-3}$. This agrees with the estimate of Venemans (\cite{bib:venemans2005}), who found a density of $\sim 6 \times 10^{-6}$~\mpc$^{-3}$ for LAE protoclusters at $z = 2 - 5.2$. They also report on similar results from Steidel et al. (\cite{bib:steidel1998}), based on LBGs at $z = 2.7 - 3.4$ ($3 \times 10^{-6}$~\mpc$^{-3}$), Shimasaku et al. (\cite{bib:shimasaku2003}), based on LAEs at $z = 4.9$ ($7 \times 10^{-6}$~\mpc$^{-3}$) and Ouchi et al. (\cite{bib:ouchi2005}), based on LAEs at $z = 5.7$ ($> 2 \times 10^{-6}$~\mpc$^{-3}$).

Our estimated volume density of protoclusters can be compared with the volume density of local rich clusters. Values found for rich cluster density at low redshift ($z < 0.1$) lie in the range $( 2 - 4 ) \times 10^{-6}$~\mpc$^{-3}$ (Bahcall \& Soneira \cite{bib:bahcall1983}; Postman et al. \cite{bib:postman1992}; Peacock \& West \cite{bib:peacock1992}; Zabludoff et al. \cite{bib:zabludoff1993}; Mazure et al. \cite{bib:mazure1996}). Our result is consistent with this number range, providing further evidence that the observed overdensities at high redshift are progenitors of rich clusters in the local Universe.

\section{Conclusions}
\label{sec:con}

We draw several conclusions from the present observations: \\
(i) TN~J1338--1942 is located in an overdensity of LBGs, both in projection and in redshift space. The new wide-field results are consistent with previous observations which revealed the presence of a protocluster through the overdensities of LAEs (Venemans et al. \cite{bib:venemans2002}; Venemans \cite{bib:venemans2005}) and LBGs (Miley et al. \cite{bib:miley2004}; Overzier et al. \cite{bib:overzier2005}). This further supports the hypothesis that HzRGs are located in dense environments. The apparent size of the overdensity and the relative position of TN~J1338--1942 within the overdensity are similar to that found by Venemans et al. for LAEs. \\
(ii) There are $4 - 7$ additional overdensities in the projected LBG distibution, similar to the one harbouring TN~J1338--1942. These may well be due to protoclusters at $z = 3.5 - 4.5$ and one or more of these overdensities could well be physically related to the TN~J1338--1942 protocluster. \\
(iii) The spatial distribution of our complete LBG sample is consistent with a Universe at $z \sim 4$ that comprises a web of compact galaxy overdensities (protoclusters) embedded in larger regions of galaxy underdensities (voids). The statistics of the overdensities are consistent with the local volume density of rich clusters.

Spectroscopic measurements are needed to investigate whether there is a physical connection between some of the outlying observed galaxy overdensities and the overdensity corresponding to the TN~J1338--1942 protocluster. Such observations could enable the cosmic web at $z = 4.1$ to be traced over distances of tens of megaparsec. Furthermore, similar measurements on other $z > 2$ protoclusters would be useful for constraining the development of large-scale structure in the early Universe.

\begin{acknowledgements}

We want to thank the anonymous referee for suggesting several improvements for this article. We are grateful to the staff of the Subaru Telescope Facility at Mauna Kea, Hawaii, for their support. We wish to recognize and acknowledge the very significant cultural role and reverence that the summit of Mauna Kea has always had within the indigenous Hawaiian community. We are most fortunate to have the opportunity to conduct observations from this mountain. HTI acknowledges a grant from the Netherlands Research School for Astronomy (NOVA). GKM acknowledges a grant from the Royal Netherlands Academy of Arts and Sciences (KNAW).

\end{acknowledgements}

\end{document}